\newcommand{\sssec}[1]{\addvspace{0.05in}\noindent\textbf{#1}}

\documentclass[manuscript,screen,nonacm]{acmart}

\usepackage{booktabs}
\usepackage{graphicx}
\usepackage[labelformat=simple]{subcaption}
\usepackage{xspace}
\usepackage{multirow}
\usepackage{enumitem}
\usepackage{microtype}
\usepackage{wasysym}
\usepackage{tabularx}
\usepackage{xcolor}
\usepackage{subcaption}
\usepackage[table]{xcolor}
\usepackage{adjustbox}
\usepackage{placeins}

\AtBeginDocument{%
  }

\setcopyright{acmlicensed}
\copyrightyear{2018}
\acmYear{2018}
\acmDOI{XXXXXXX.XXXXXXX}
\acmConference[Conference acronym 'XX]{Make sure to enter the correct
  conference title from your rights confirmation email}{June 03--05,
  2018}{Woodstock, NY}
\acmISBN{978-1-4503-XXXX-X/2018/06}

\begin{document}

\title{Auditing Hyperstimulation at Scale in Child-Facing YouTube Videos}

\author{Shao-Yu Chu}
\affiliation{%
  \institution{University of California, San Diego}
  \city{La Jolla}
  \country{United States}}
\email{shaoyuchu@ucsd.edu}

\author{Qiyu Li}
\affiliation{%
  \institution{University of California, San Diego}
  \city{La Jolla}
  \country{United States}}
\email{qiyuli@ucsd.edu}

\author{Xuan Liu}
\affiliation{%
  \institution{University of California, San Diego}
  \city{La Jolla}
  \country{United States}}
\email{xul049@ucsd.edu}

\author{Shuang Hao}
\affiliation{%
  \institution{University of Texas at Dallas}
  \city{Dallas}
  \country{United States}}
\email{shao@utdallas.edu}

\author{Haojian Jin}
\affiliation{%
  \institution{University of California, San Diego}
  \city{La Jolla}
  \country{United States}}
\email{haojian@ucsd.edu}

\renewcommand{\shortauthors}{Chu et al.}

\begin{abstract}
Parents, pediatricians, and journalists have raised concern that child-facing videos are becoming increasingly stimulating (faster-paced, louder, and more visually dense), but this concern rests mostly on ad hoc observation rather than systematic measurement. We present a large-scale audit of hyperstimulation patterns, which we refer to as content-independent production choices, such as rapid cutting, that may place demands on children's developing cognitive and sensory processing capacities. Drawing on parents' Reddit discussions and developmental psychology, we develop computational metrics for 12 hyperstimulation patterns. We apply these metrics to 16,375 videos from 542 child-facing YouTube channels spanning 2015--2026. We found broad increases in stimulation over time, with cut rates rising across all five content categories. Nearly a quarter of the channels increased in at least one stimulation pattern over time. More stimulating videos also tended to receive more views, but the relevant stimulation patterns differed across content types.

\end{abstract}



\maketitle

\section{Introduction}

Parents, pediatricians, and journalists have raised growing concern about the increasing stimulation caused by the bright colors and rapid scene changes in child-facing videos~\cite{radesky-2020-youtube, AAP-2016, Whatisth72:online}. Parents report that children who watch CoComelon, one of the most-viewed channels on YouTube, become distressed and agitated when viewing is interrupted~\cite{lloyd-2023}. Developmental psychologists also found that just nine minutes of a fast-paced cartoon was enough to impair 4-year-olds' executive function compared to educational programming~\cite{lillard-2011}. However, discussion of these stimulations in child-facing videos is mostly based on ad hoc observations; our society has not yet systematically characterized stimulation across children’s media at scale.

In this paper, we conduct a large-scale computational audit of hyperstimulation patterns, which we define as content-independent production choices (such as rapid cutting, sustained loudness, or the elimination of silence) that may push videos beyond what young children can comfortably attend to~\cite{anderson-lorch-1983}, process~\cite{lang-2000}, or regulate~\cite{lillard-2011}. We use the term \textit{hyperstimulation}~\cite{ofcom2025childrens} rather than \textit{overstimulation}~\cite{christakis2004early, Cocomelo76:online} because we restrict our claims to measurement: we document which hyperstimulation patterns exist and how prevalent they are, not whether they harm children. We hope this audit helps parents make informed choices when selecting videos for their children, helps platforms surface child-safety signals, and helps creators make more accountable decisions about the stimulation levels of their videos. We make no claims about whether such stimulation causes addiction, withdrawal, or developmental harm.

\begin{figure}[t]
    \centering
    \includegraphics[width=\linewidth]{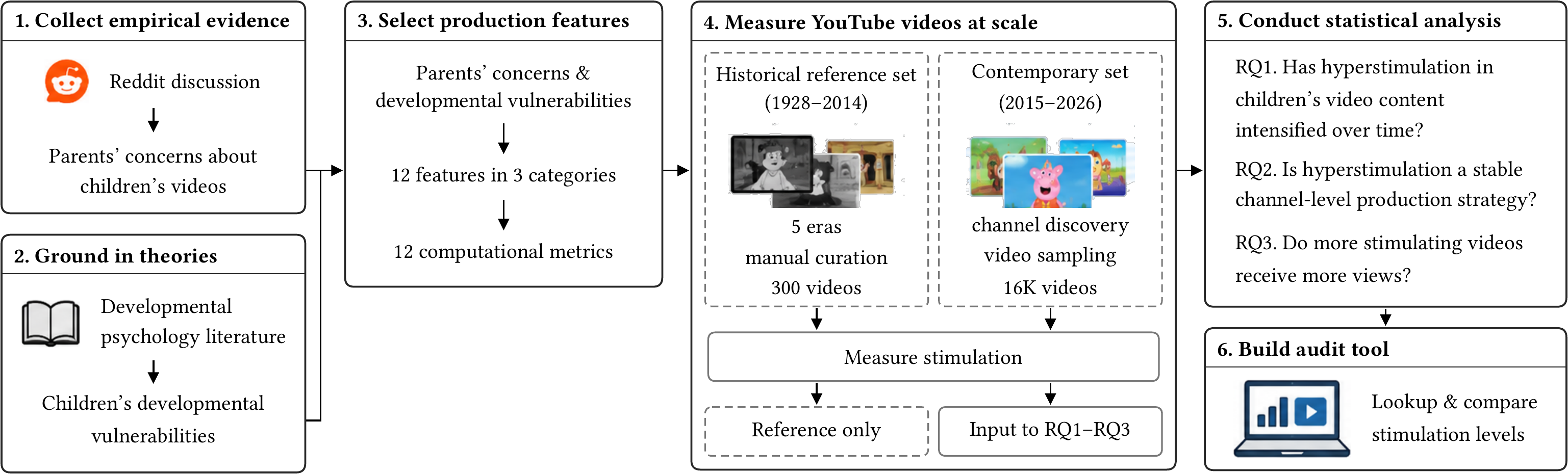}
    \caption{Research overview. We identified 12 stimulation patterns based on parents’ concerns and developmental vulnerabilities documented in the psychology literature. We measure these patterns across 16K children’s YouTube videos and build a public audit tool for parents, content creators, and regulators to assess videos’ stimulation levels.}
    \label{fig:research-methods}
    \vspace{-0.1in}
\end{figure}

Figure~\ref{fig:research-methods} illustrates our measurement framework.
We first analyzed 155 Reddit posts and 811 comments to identify concerning audiovisual production choices. Building on these concerns and developmental psychology literature, we selected 12 hyperstimulation patterns organized into three categories: Pacing Intensification, Sustained Sensory Intensity, and Reward Patterning (Section~\ref{understanding-hyperstimulation}). We operationalized each pattern as a computational metric (Section~\ref{measure-hyperstimulation}) and applied the framework to 16,375 videos from 542 child-facing YouTube channels released between 2015--2026, with 300 historical videos (1928–2014) serving as reference points. We conducted analyses to answer the following research questions (RQs):

\begin{itemize}[nosep,leftmargin=*]
    \item \textbf{RQ1: Have hyperstimulation patterns intensified over time?} Hyperstimulation increased across multiple production choices, with rapid cutting rising across all five content categories. The most extreme stimulation levels increased while lower levels remained relatively stable. For example, among gaming videos, cut rates at the 95th percentile increased by 1.45 cuts/min per year, while those at the 5th percentile remained near zero.

    \item \textbf{RQ2: Are stimulation levels consistent within channels?} 
    Stimulation levels were not consistent within channels. Nearly a quarter of channels increased in at least one stimulation pattern over time. For example, 7.4\% of channels significantly increased their cut rates, the most widespread increase among the 12 patterns.

    \item \textbf{RQ3: Do more stimulating videos receive more views?} 
    Higher stimulation levels were associated with more views, but the relevant stimulation patterns differed across content categories. More views were associated with faster cutting in narrative videos, greater audio stimulation in gaming and music videos, and greater visual stimulation in educational and play \& activity videos.
\end{itemize}

We also developed a public tool that lets parents, regulators, and creators look up the stimulation levels of YouTube videos (Section~\ref{sec:audit-tool}). We will release the tool as open source alongside our dataset and measurement pipeline. This paper makes the following contributions: 
\begin{enumerate}[nosep,leftmargin=*]
    \item Our large-scale audit finds increasing hyperstimulation across child-facing YouTube, with nearly a quarter of channels showing increases in at least one pattern. Higher stimulation levels are also associated with more views, with patterns varying across content categories.
    \item We develop an open-source pipeline to measure 12 hyperstimulation patterns informed by parents' concerns and children's developmental vulnerabilities, compile a dataset of 16,375 contemporary children's YouTube videos (2015–2026), and build a public tool for looking up videos' stimulation levels.
\end{enumerate}

\section{Related Work}

Our work draws on three bodies of research: the developmental psychology literature on media pacing and children's cognition, the HCI literature on manipulative design, and computational approaches to analyzing children's video.
\sssec{Media Pacing, Formal Features, and Children's Cognition}. 
Prior work shows that production features such as cuts, pacing, and sound shape how children attend to and process video. Huston and Wright~\cite{huston-wright-1983} termed these \emph{formal features} and showed that they elicit orienting responses in young children; Anderson et al.~\cite{anderson-lorch-1983, anderson-pempek-2005} found that movement and cuts sustain looking in preschoolers. Lang's Limited Capacity Model~\cite{lang-2000} explains why this can backfire: each cut prompts an orienting response that aids encoding, but excessive stimulation overwhelms processing capacity and impairs comprehension, particularly in young children. Lillard and Peterson~\cite{lillard-2011} showed that nine minutes of a fast-paced cartoon impaired 4-year-olds' executive function relative to educational programming, and longitudinal work by Christakis and colleagues~\cite{christakis2004early, zimmerman-2007, christakis-2019} linked early entertainment television exposure to later attention problems, but not educational television exposure, suggesting that production features matter beyond exposure quantity.

However, less is known about how these features appear and change in child-facing media. Most existing experimental literature characterizes today's media by assertion: Lillard and Peterson~\cite{lillard-2011} motivate their study by noting that contemporary programs are faster-paced than those of earlier decades, and Christakis~\cite{christakis-2019} describes modern content as increasingly stimulating, but both rely on illustrative examples rather than systematic measurement.
We address this gap by computationally measuring the prevalence and evolution of production features at scale.

\sssec{Persuasive Design and Manipulative Engagement}. 
The concern that media design can influence audiences without their awareness has a long intellectual history. Packard's \emph{The Hidden Persuaders}~\cite{packard-1957} documented how the advertising industry adopted techniques that improved metrics without understanding the psychological mechanisms at work. Fogg~\cite{fogg-2003} formalized this for digital technology as \emph{persuasive design}. More recently, literature on dark patterns has shown that manipulative UI choices are widespread in commercial platforms~\cite{gray2018dark, mathur-2019, narayanan-2020, mathur2021makes}. Researchers have also developed computational approaches to detecting these patterns, including the use of multimodal language models to identify deceptive interface designs~\cite{kocyigit2025deceptilens}. Meanwhile, work on engagement mechanisms shows that variable reinforcement schedules and reward-learning principles drive habit formation in digital products~\cite{lindstrom-2021, eyal2014hooked}. However, recent work cautions that whether individual design patterns are fair depends on their broader context~\cite{de2025if}.

When applied to children, these dynamics are particularly concerning. Radesky et al.~\cite{radesky2022prevalence} found a majority of apps used by 3--5-year-old children associates with manipulative design features. Recent work~\cite{krahl2025playful} showed that in children's apps, engagement mechanisms are interwoven with core interaction mechanics and sensory cues. However, this literature focuses on interactive interfaces, leaving open whether video production choices similarly optimize engagement without regard for developmental impact.

In contrast, our work applies the Packard--Fogg--Mathur lineage~\cite{packard-1957,fogg-2003,mathur-2019, mathur2021makes} to the production of child-facing video. While dark patterns research examines deliberate UI deception, we examine production choices that may emerge from metric optimization. 
Our measurement framework makes these production patterns visible and quantifiable.

\sssec{Computational Audits of YouTube Videos}. 
Prior computational audits of YouTube have examined potential platform harms. Ribeiro et al.~\cite{ribeiro2020auditing} analyzed over 330K videos to investigate radicalization pathways, while Nigatu and Raji~\cite{nigatu2024searched} combined interviews with analyses of YouTube search results and recommendations to identify harmful content exposure in low-resourced languages. Research focusing on children's YouTube has primarily examined content safety and learning value. Researchers have built classifiers to detect age-inappropriate content, such as sexual and violent content, using video metadata~\cite{papadamou-2020} and audiovisual features~\cite{kidstube-2019, kaushal2016kidstube, alghowinem2018safer, han2020discovery}. Others quantify the educational value of children's videos~\cite{kumar2026quantifying}. These studies examine video content rather than audiovisual stimulation.

Recently, Söğütlüler~\cite{sougutluler10hyperstimulation} proposed a computational hyperstimulation index for children's media and applied it to 90 top-viewed videos from 30 YouTube channels. Similarly, the commercial directory TV Tantrum~\cite{tvtantrum} rates around 300 children's TV shows by their stimulation levels. However, both provide snapshots of selected content. In contrast, we examine changes over time and within channels through a large-scale, reproducible audit of more than 16K children's YouTube videos spanning 12 years, with an open-source pipeline for computing all stimulation metrics.

\section{Child-Facing Video Dataset}

We constructed a contemporary YouTube corpus spanning 2015--2026 to examine longitudinal changes in hyperstimulation (RQ1), within-channel consistency (RQ2), and associations with video popularity (RQ3). We also curated a historical reference set (1928--2014) to provide descriptive context for stimulation levels in earlier children's videos.

\begin{figure}[h]
  \centering
  \includegraphics[width=\textwidth]{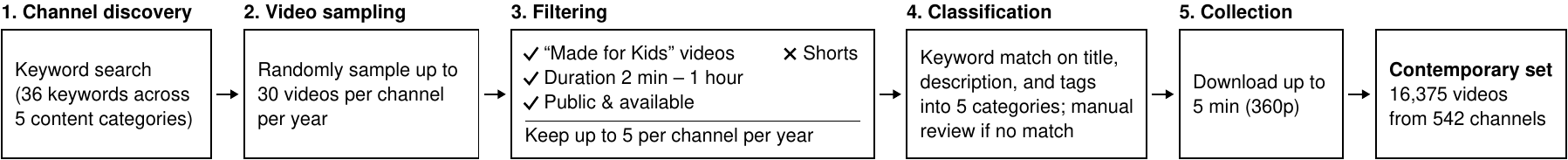}
  \vspace{-15pt}
  \caption{Contemporary dataset construction workflow.}
  \label{fig:dataset-workflow}
\end{figure}

\sssec{Contemporary YouTube Corpus}.
We used a channel-first search. We first identified channels that produce children’s content and then sampled videos from each channel. This approach prevents the corpus from being dominated by a small number of channels, as can occur when treating videos as the search unit. Figure~\ref{fig:dataset-workflow} summarizes the construction of the contemporary corpus.

\vspace{-8pt}

\sssec{Channel Discovery}.
We targeted five content categories representing the dominant genres of child-facing YouTube content: \textit{Music}, \textit{Narrative}, \textit{Educational}, \textit{Gaming}, and \textit{Play \& Activity}. We searched for channels using 36 keywords (Appendix~\ref{app:search-queries}) via the YouTube Data API. We required channels to have (1)~at least 10K subscribers, as YouTube viewership is highly skewed, with a small share of popular channels accounting for a large share of views~\cite{cha2007tube}, and (2)~at least 100 videos, ensuring sufficient upload history for the channel-level analyses in RQ2. We excluded ``Topic'' channels, since YouTube auto-generates them to populate its music-streaming service, rather than to host video content that a creator intentionally produced for the video platform~\cite{youtube_find_music_artists}. 
This process yielded 646 channels.

\sssec{Video Sampling}.
We sampled videos across multiple years per channel to capture temporal variation in production style and avoid overrepresenting recently uploaded content. For each channel, we first randomly sampled up to 30 videos per year from 2015 onward. We then excluded videos that were (1) shorter than 2 minutes (often trailers for upcoming episodes), or longer than 1 hour (often compilations of existing content); (2) not labeled as ``Made for kids’’\cite{youtube_made_for_kids} by their creator; (3) private or unavailable in our region; or (4) in \textit{Shorts} format. Shorts are vertical, short-form videos that are often repurposed from existing long-form videos~\cite{barua2025lotus}, which naturally results in higher cut rates. From the remaining videos, we randomly sampled at most 5 videos per year per channel, ensuring that the sample captures changes over time for RQ1 without highly active channels disproportionately dominating the dataset. Eighteen videos were no longer available during the download. The final corpus yields 16,375 videos from 542 channels.

\sssec{Video classification.}
We classified each contemporary video into one of five content categories by matching its metadata (title, description, tags) against category-specific keyword lists. Videos matching multiple categories were assigned by priority order: \textit{gaming}, \textit{music}, \textit{narrative}, \textit{play \& activity}, \textit{educational}. Authors manually reviewed videos matching none of the keyword lists and manually assigned their categories. Appendix~\ref{app:contemporary-distribution} presents the distribution.

\sssec{Historical Reference Set}.
We curated 300 historical children’s videos across five production eras (1928--2014), including at most three videos per series per era. We excluded videos with uncertain production years or severe audiovisual degradation. Appendix~\ref{app:historical-dataset} details the collection procedure. Because production technologies and remastering can affect measured stimulation levels, we used this set only as descriptive reference points in figures, excluding it from RQ1--RQ3.

\sssec{Data Collection}.
We downloaded up to five minutes of each video at 360p resolution. Contemporary clips began at 30 seconds to skip intros, while historical clips began at an author-selected point to accommodate longer openings. We collected view counts, upload dates, titles, descriptions, and tags for contemporary videos, but only titles for historical videos, as other metadata reflects re-uploads rather than original content. We collected all videos in August 2026.

\vspace{-4pt}
\section{Computational Audit of Hyperstimulation in Child-Facing Videos}

We develop a computational framework for measuring hyperstimulation in child-facing videos. We first select production features informed by parents' concerns and developmental psychology (Section~\ref{understanding-hyperstimulation}), then operationalize them as computational metrics (Section~\ref{measure-hyperstimulation}).

\subsection{Understanding Hyperstimulation}
\label{understanding-hyperstimulation}

\sssec{Parents' Concerns.} We collected 155 posts and 811 comments (Jul. 2024--Dec. 2025) discussing hyperstimulation in children’s videos across eight parenting-related subreddits (Appendix~\ref{app:subreddits}). Following dual screening for relevance, two authors developed a codebook through iterative open coding and coded the discussions to identify recurring concerns. We detail the screening and coding procedures in Appendix~\ref{app:coding}.

Parents' concerns centered on rapid pacing and sustained sensory intensity. One parent described shows with ``lots of camera cuts'' as ``a bit overwhelming for their developing visual and attention systems.'' Others highlighted bright, saturated colors, loud audio, and the absence of quiet moments. One described videos featuring ``bright colors and loud sounds'' as ``dopamine machines.'' Parents also worried that children accustomed to highly stimulating videos might lose interest in slower content. As one observed, ``when we put on one of the slower shows, he's bored within seconds.''

\sssec{Developmental Vulnerabilities.} To understand the developmental mechanisms underlying these concerns and explore whether other production characteristics might operate similarly, we turned to developmental psychology research. We focused on vulnerabilities that (1) have empirical support, (2) relate to audiovisual production rather than narrative content, and (3) are relevant to children under 13. We identified two vulnerabilities. 
\textbf{Cognitive depletion} concerns children's limited cognitive resources for attention and self-regulation. Research on children's television viewing shows that formal features, such as cuts, movement, and sound effects, can elicit orienting responses and sustain attention~\cite{anderson-lorch-1983, huston-wright-1983}. Each change redirects attention and requires cognitive resources to process new information. When changes occur frequently, their combined demands may exceed children's available processing capacity~\cite{lang-2000}. 
\textbf{Sensory calibration} concerns how early experiences shape developing arousal and reward systems. Early childhood is a period of heightened neuroplasticity~\cite{darling2020first, sahinoglu2026neonatal}, during which sensory experiences interact with arousal regulation~\cite{jones2003arousal, robbins1997arousal} and reward processing~\cite{arias2010dopaminergic}. Repeated exposure to intense sensory input may alter what children become accustomed to, potentially making less stimulating experiences less engaging.

\sssec{Selecting Hyperstimulation Patterns.} Guided by parents' concerns and the developmental vulnerabilities described above, we selected 12 hyperstimulation patterns that capture potentially demanding audiovisual production choices. We first translated recurring concerns into measurable features, such as rapid cutting and sustained loudness, then extended them based on developmental mechanisms. For example, cognitive depletion motivated us to examine not only cut frequency but also visual discontinuity between shots, which may increase processing demands. We prioritized patterns that (1) relate to the identified developmental vulnerabilities, (2) capture distinct production characteristics, and (3) can be measured from audiovisual signals without semantic interpretation. For example, we excluded average shot duration because it largely overlaps with Rapid Cutting. We also excluded audio-content characteristics such as singing and background music presence, as their presence alone does not indicate greater cognitive processing demands or sustained sensory intensity.

\begin{figure}[h]
    \centering
    \includegraphics[width=\linewidth]{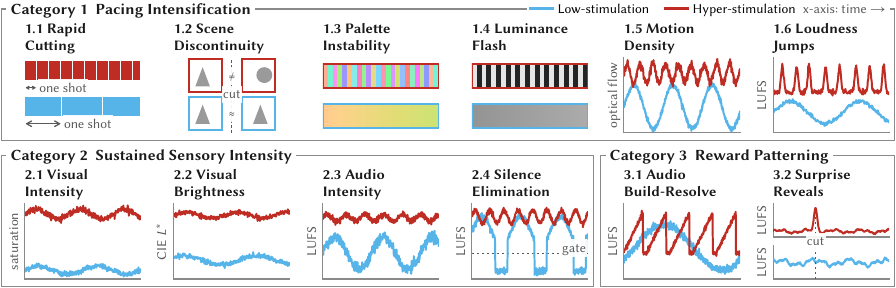}
    \vspace{-15pt}
    \caption{Illustrations of the 12 hyperstimulation patterns, contrasting lower- and higher-stimulation audiovisual signals.}
    \label{fig:stimulation-patterns}
\end{figure}

We organized the 12 patterns into three categories: \textbf{Pacing Intensification} captures frequent audiovisual changes that may tax children's limited cognitive resources; \textbf{Sustained Sensory Intensity} captures prolonged high-intensity stimulation; and \textbf{Reward Patterning} captures audiovisual intensity changes that produce distinct sensory peaks. The latter two relate to sensory calibration through arousal and reward processing. Table~\ref{tab:production-choice} summarizes the patterns and their developmental motivations, and Figure~\ref{fig:stimulation-patterns} illustrates their audiovisual characteristics.

\begin{table}[t]
\centering
\footnotesize
\caption{Hyperstimulation patterns. P = Parents' Concerns,
T = Theory-Informed Extension; CD = Cognitive Depletion,
SC = Sensory Calibration; V = Visual, A = Audio.}
\vspace{-5pt}
\label{tab:production-choice}
\begin{tabular}{@{}l p{6.5cm} cc cc cc@{}}
\toprule
\textbf{Pattern} & \textbf{Definition}
& \multicolumn{2}{c}{\textbf{Origin}}
& \multicolumn{2}{c}{\textbf{Vulnerability}}
& \multicolumn{2}{c}{\textbf{Modality}} \\
\cmidrule(lr){3-4}
\cmidrule(lr){5-6}
\cmidrule(lr){7-8}
& & P & T & CD & SC & V & A \\
\midrule
\multicolumn{8}{@{}p{0.95\linewidth}}{\textbf{Pacing Intensification}} \\[2pt]
(1.1) Rapid Cutting
& A high rate of shot changes.
& \checkmark & & \checkmark & & \checkmark & \\
(1.2) Scene Discontinuity
& Cuts where pre- and post-cut frames share low visual similarity.
& & \checkmark & \checkmark & & \checkmark & \\
(1.3) Palette Instability
& Rapid color change between frames.
& & \checkmark & \checkmark & & \checkmark & \\
(1.4) Luminance Flash
& Rapid brightness change between frames.
& \checkmark & & \checkmark & & \checkmark & \\
(1.5) Motion Density
& Sustained optical flow that rarely settles into visual stillness.
& \checkmark & & \checkmark & & \checkmark & \\
(1.6) Loudness Jumps
& Frequent jumps in loudness.
& & \checkmark & \checkmark & & & \checkmark \\
\midrule
\multicolumn{8}{@{}p{0.95\linewidth}}{\textbf{Sustained Sensory Intensity}} \\[2pt]
(2.1) Visual Intensity
& Sustained high color saturation.
& \checkmark & & & \checkmark & \checkmark & \\
(2.2) Visual Brightness
& Sustained high luminance.
& \checkmark & & & \checkmark & \checkmark & \\
(2.3) Audio Intensity
& Sustained high loudness.
& \checkmark & & & \checkmark & & \checkmark \\
(2.4) Silence Elimination
& Low proportion of quiet moments.
& \checkmark & & & \checkmark & & \checkmark \\
\midrule
\multicolumn{8}{@{}p{0.95\linewidth}}{\textbf{Reward Patterning}} \\[2pt]
(3.1) Audio Build-Resolve
& Short cycles of rising loudness followed by a sharp drop.
& & \checkmark & & \checkmark & & \checkmark \\
(3.2) Surprise Reveals
& Visual cut co-occurs with an audio energy spike.
& & \checkmark & & \checkmark & \checkmark & \checkmark \\
\bottomrule
\end{tabular}
\end{table}

\subsection{Measuring Hyperstimulation}
\label{measure-hyperstimulation}
We operationalize the 12 hyperstimulation patterns as computational metrics~\cite{jacobs2021measurement}, prioritizing perceptually grounded measures over raw audiovisual signals. We focus on four underlying properties--cuts, colors, motion, and loudness--to derive one metric per pattern, with higher values indicating greater stimulation. We present the metrics in Table~\ref{tab:metrics}.

\begin{table}[h]
\centering
\caption{Computational metrics for the 12 hyperstimulation patterns. Higher values indicate greater stimulation.}
\vspace{-5pt}
\Description{Operationalization of 12 hyperstimulation patterns across three categories: Pacing Intensification, Sustained Sensory Intensity, and Reward Patterning. Each pattern is mapped to a computational metric, with higher values indicating greater stimulation.}
\footnotesize
\begin{tabular}{@{}
  >{\raggedright\arraybackslash}p{0.17\linewidth}
  >{\raggedright\arraybackslash}p{0.12\linewidth}
  >{\raggedright\arraybackslash}p{0.62\linewidth}
@{}}
\toprule
\textbf{Pattern} & \textbf{Metric} & \textbf{Operationalization} \\
\midrule
\multicolumn{3}{@{}l}{\textbf{Pacing Intensification}} \\
\addlinespace[2pt]
1.1 Rapid Cutting
& cuts/min
& Detected shot boundaries per minute (\texttt{scdet}, threshold 11). \\

1.2 Scene Discontinuity
& mean DSSIM
& Mean structural dissimilarity ($1-\mathrm{SSIM}$) across cut boundaries. \\

1.3 Palette Instability
& changes/min
& One-second CIELAB palette changes with Jaccard distance $>0.45$, per minute. \\

1.4 Luminance Flash
& flashes/min
& Increases $\geq 20$ in mean $L^*$ between adjacent 15\,fps samples within a shot, per minute. \\

1.5 Motion Density
& motion fraction
& Share of 15\,fps intervals with mean optical flow $\geq 10\%$ of frame width/s. \\

1.6 Loudness Jumps
& jumps/min
& Moments $\geq 6$\,LU above integrated loudness, at most once per second. \\

\midrule
\multicolumn{3}{@{}l}{\textbf{Sustained Sensory Intensity}} \\
\addlinespace[2pt]
2.1 Visual Intensity
& mean saturation
& Mean HSV saturation across pixels. \\

2.2 Visual Brightness
& mean $L^*$
& Mean CIELAB lightness across pixels. \\

2.3 Audio Intensity
& LUFS
& Integrated loudness (EBU R128, K-weighted). \\

2.4 Silence Elimination
& active fraction
& Share of 100\,ms blocks above the higher of mean energy loudness $-10$\,LU or $-70$\,LUFS. \\

\midrule
\multicolumn{3}{@{}l}{\textbf{Reward Patterning}} \\
\addlinespace[2pt]
3.1 Audio Build-Resolve
& cycles/min
& Loudness rises over 1.2--5\,s followed by a 45--130\% drop within 1\,s, per minute. \\

3.2 Surprise Reveals
& events/min
& Loudness jumps occurring within 0.2\,s of a cut, per minute. \\

\bottomrule
\end{tabular}
\label{tab:metrics}
\end{table}

\sssec{Cuts.} We detect cuts by identifying abrupt visual changes that exceed normal frame-to-frame variation. We then measure discontinuity at each cut using structural dissimilarity ($1 - \text{SSIM}$). Unlike raw pixel differences, SSIM~\cite{wang2004image} accounts for luminance, contrast, and structure to better approximate perceived visual differences.

\begin{figure}[h]
    \centering
    \includegraphics[width=0.9436\linewidth]{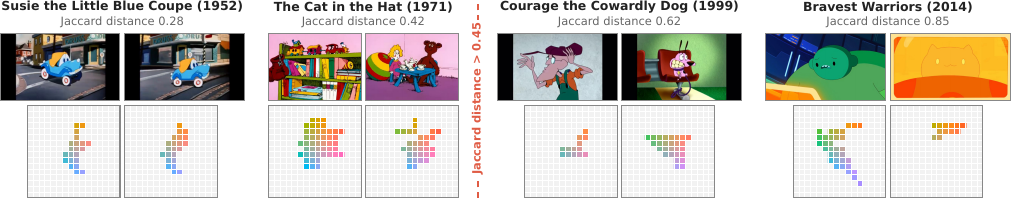}
    \vspace{-5pt}
    \caption{We quantify \textbf{palette change} as the Jaccard distance between two frames' sets of occupied CIELAB a*b* tiles, computed over a 16×16 grid spanning 16 Lab units per tile. A tile counts as presented when it holds $\geq 0.1$\% of the frame's pixels. We count two frame changes on the right as palette changes since their Jaccard distance is above the 0.45 threshold. 
    }
    \label{fig:palette-instability-explainer}
\end{figure}

\begin{figure}[h]
    \centering
    \includegraphics[width=0.8447\linewidth]{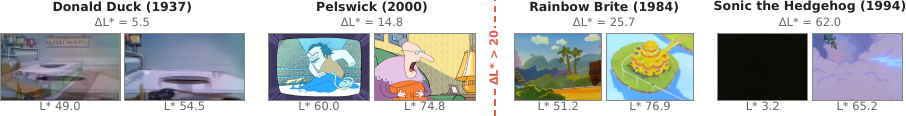}
    \vspace{-5pt}
    \caption{We measure \textbf{luminance} as the mean L* value in CIELAB space across pixels. A \textit{luminance flash} occurs when the luminance difference between adjacent frames, sampled at 15 fps, exceeds 20.}
    \label{fig:luminance-flash-explainer}
\end{figure}

\sssec{Colors.} We primarily use CIELAB~\cite{robertson1977cie} because numerical differences in this space better approximate human-perceived color differences than in other color spaces, such as RGB. We use its chromatic axes (a*, b*) to measure palette changes (Figure~\ref{fig:palette-instability-explainer}) and its lightness axis (L*) to measure luminance flashes (Figure~\ref{fig:luminance-flash-explainer}) and visual brightness. For visual intensity, we instead use HSV saturation, which captures color vividness relative to gray.

\begin{figure}[h]
    \centering
    \includegraphics[width=0.9884\linewidth]{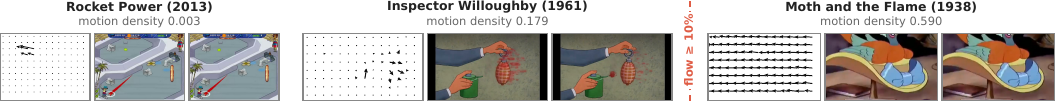}
    \vspace{-5pt}
    \caption{We detect \textbf{motion} when mean optical flow magnitude exceeds 10\% of frame width per second.}
    \label{fig:motion-density-explainer}
\end{figure}

\sssec{Motion.} We use optical flow to measure frame-to-frame visual movement, capturing both object and camera motion without relying on semantic content (Figure~\ref{fig:motion-density-explainer}).

\sssec{Loudness.} We use LUFS~\cite{series2011algorithms} as it accounts for human sensitivity to different audio frequencies, better approximating perceived loudness than raw amplitude. Unlike psychoacoustic measures such as sones~\cite{stevens1955measurement}, which depend on playback conditions, LUFS provides a consistent basis for comparing audio intensity and its temporal changes across videos.

\sssec{Implementation.}
We build the measurement pipeline using open-source FFmpeg and OpenCV to support accessible and reproducible auditing. Rather than decoding each video separately for all 12 metrics, we group computations into three passes: frame-to-frame changes (cuts, flashes, and motion), color analysis, and audio analysis. For the cross-pass metric (3.2) Surprise Reveals, we align loudness jumps with cut timestamps.

\section{Computational Audit at Scale}

We apply the 12 hyperstimulation metrics to our contemporary video set to examine longitudinal change (Section~\ref{rq1}), within-channel consistency (Section~\ref{rq2}), and associations with video popularity (Section~\ref{rq3}).

\subsection{RQ1: Have hyperstimulation patterns intensified over time?}
\label{rq1}

\subsubsection{Method}

To examine changes in stimulation over time across content categories, we fitted two families of models according to their metric distributions. For Pacing Intensification (1.1--1.6) and Sustained Sensory Intensity (2.1--2.4) metrics, we fitted quantile regressions~\cite{koenker2001quantile} at the 5th, 50th, and 95th percentiles:
\[
Q_\tau(Y_m \mid \mathrm{Year}, \mathrm{Category})
= \beta_0^\tau
+ \beta_1^\tau \cdot \mathrm{Year}
+ \beta_2^\tau \cdot \mathrm{Category}
+ \beta_3^\tau \cdot (\mathrm{Year} \times \mathrm{Category}),
\]
where \(\tau \in \{0.05, 0.50, 0.95\}\) denotes the quantile level, and \(Y_m\) denotes the raw value of stimulation metric \(m\). The model includes upload year, content category, and their interaction. We chose quantile regression over linear regression for two reasons. First, quantile regression captures changes at the lower, median, and upper ends of the stimulation distribution, rather than only the mean. Second, seven of the ten metrics exhibited heteroscedasticity, which violates the constant-variance assumption of ordinary least squares regression. We estimated category-specific annual slopes, treating 95\% confidence intervals excluding zero as significant. We also tested Year $\times$ Category interactions to assess whether trends differed across categories, applying Benjamini–Hochberg (BH) correction~\cite{benjamini1995controlling} to the interaction tests.

For the two Reward Patterning metrics (3.1, 3.2), more than half of the observations were zero, making quantile regression unsuitable for characterizing trends among videos exhibiting these patterns. We therefore fitted two-part hurdle mixed models. The first part used logistic regression to estimate annual changes in the odds of a pattern occurring, while the second used a zero-truncated count model to estimate changes in its frequency among videos exhibiting the pattern. Both included upload year, content category, and their interaction, with BH correction applied to hypothesis tests. Appendix~\ref{app:rq1-methods} details model specifications and diagnostics.

\subsubsection{Results.}
Across the 50 metric--category combinations analyzed using quantile regression (10 metrics $\times$ 5 content categories), 32 (64\%) showed a significant increase at one or more percentiles, while 12 (24\%) showed significant decreases but no increases and six (12\%) showed no significant changes (Table~\ref{tab:longitudinal-slope}).

\newlength{\metriccolwidth}
\setlength{\metriccolwidth}{0.09\linewidth}

\definecolor{salmoncolor}{RGB}{255,160,122}
\definecolor{tealcolor}{RGB}{102,178,178}

\newcommand{\Nc}[1]{$#1$}
\newcommand{\Sc}[1]{\cellcolor{salmoncolor!45}$\mathbf{#1}$}
\newcommand{\Tc}[1]{\cellcolor{tealcolor!45}$\mathbf{#1}$}

\newcommand{\Up}{\cellcolor{salmoncolor!45}$\uparrow$}
\newcommand{\Dn}{\cellcolor{tealcolor!45}$\downarrow$}
\newcommand{\Ns}{$-$}

\begin{table}[h]
\centering
\footnotesize
\caption{Annual slope estimates for the ten stimulation
metrics analyzed using quantile regression. Each category
reports the 95th, 50th, and 5th percentiles. Shaded cells
indicate slopes whose 95\% confidence intervals exclude
zero: \colorbox{salmoncolor!45}{salmon} for increases and
\colorbox{tealcolor!45}{teal} for decreases.
Full estimates, confidence intervals, and interaction
tests are reported in Appendix~\ref{app:model-estimates-rq1}.}
\renewcommand{\arraystretch}{1.2}
\setlength{\tabcolsep}{2pt}

\begin{adjustbox}{max width=\linewidth}
\begin{tabular}{@{}l *{5}{ccc}@{}}
\toprule
& \multicolumn{3}{c}{Music}
& \multicolumn{3}{c}{Narrative}
& \multicolumn{3}{c}{Educational}
& \multicolumn{3}{c}{Gaming}
& \multicolumn{3}{c}{Play \& Activity} \\
\cmidrule(lr){2-4}
\cmidrule(lr){5-7}
\cmidrule(lr){8-10}
\cmidrule(lr){11-13}
\cmidrule(l){14-16}
Metric
& p95 & p50 & p5
& p95 & p50 & p5
& p95 & p50 & p5
& p95 & p50 & p5
& p95 & p50 & p5 \\
\midrule

1.1 Rapid Cutting
& \Sc{0.1550} & \Sc{0.4026} & \Sc{0.0211}
& \Nc{0.1330} & \Sc{0.2389} & \Nc{0.0000}
& \Sc{0.5204} & \Sc{0.2316} & \Nc{0.0000}
& \Sc{1.4541} & \Sc{0.5003} & \Nc{0.0000}
& \Sc{0.6898} & \Nc{-0.0075} & \Nc{0.0000} \\

1.2 Scene Discontinuity
& \Tc{-0.0026} & \Sc{0.0036} & \Sc{0.0104}
& \Nc{-0.0008} & \Sc{0.0013} & \Nc{-0.0006}
& \Nc{-0.0015} & \Sc{0.0047} & \Nc{0.0024}
& \Nc{0.0010} & \Sc{0.0079} & \Nc{0.0045}
& \Nc{-0.0010} & \Tc{-0.0031} & \Nc{-0.0036} \\

1.3 Palette Instability
& \Tc{-0.3485} & \Tc{-0.1165} & \Nc{-0.0067}
& \Nc{0.0000} & \Sc{0.1040} & \Nc{0.0000}
& \Sc{0.2880} & \Nc{0.0872} & \Nc{0.0000}
& \Sc{0.7091} & \Sc{0.3874} & \Nc{0.0339}
& \Nc{0.0051} & \Nc{0.0256} & \Nc{0.0000} \\

1.4 Luminance Flash
& \Tc{-0.0358} & \Nc{0.0000} & \Nc{0.0000}
& \Nc{-0.0003} & \Nc{0.0000} & \Nc{0.0000}
& \Nc{-0.0027} & \Nc{0.0000} & \Nc{0.0000}
& \Sc{0.0845} & \Nc{0.0000} & \Nc{0.0000}
& \Nc{-0.0194} & \Nc{0.0000} & \Nc{0.0000} \\

1.5 Motion Density
& \Nc{-0.0032} & \Sc{0.0061} & \Sc{0.0020}
& \Nc{0.0035} & \Sc{0.0026} & \Nc{0.0000}
& \Nc{-0.0042} & \Sc{0.0039} & \Nc{-0.0003}
& \Tc{-0.0238} & \Nc{0.0020} & \Nc{0.0007}
& \Tc{-0.0064} & \Tc{-0.0079} & \Tc{-0.0008} \\

1.6 Loudness Jumps
& \Tc{-0.0844} & \Nc{0.0000} & \Nc{0.0000}
& \Nc{0.0001} & \Sc{0.0617} & \Nc{0.0000}
& \Tc{-0.1056} & \Tc{-0.0731} & \Nc{0.0000}
& \Nc{-0.0311} & \Tc{-0.1166} & \Nc{0.0000}
& \Nc{-0.0372} & \Tc{-0.0515} & \Nc{0.0000} \\

\midrule

2.1 Visual Intensity
& \Tc{-1.3156} & \Tc{-0.5351} & \Sc{1.7950}
& \Tc{-1.6730} & \Tc{-0.6875} & \Nc{-0.1246}
& \Sc{1.8950} & \Sc{1.2948} & \Nc{-0.6319}
& \Sc{1.7204} & \Nc{0.3371} & \Nc{-1.3163}
& \Nc{0.4196} & \Nc{-0.0916} & \Nc{-0.2823} \\

2.2 Visual Brightness
& \Tc{-0.6091} & \Sc{0.1548} & \Sc{0.6121}
& \Sc{0.4192} & \Sc{0.1888} & \Nc{0.1006}
& \Nc{0.1533} & \Nc{0.0395} & \Sc{0.5738}
& \Sc{0.7538} & \Sc{0.8744} & \Nc{0.5422}
& \Sc{0.4198} & \Nc{0.1830} & \Tc{-0.5645} \\

2.3 Audio Intensity
& \Tc{-0.1996} & \Tc{-0.1333} & \Nc{0.0047}
& \Tc{-0.3693} & \Tc{-0.1426} & \Nc{0.0560}
& \Nc{0.0674} & \Sc{0.2775} & \Sc{0.2675}
& \Nc{0.0256} & \Sc{0.2545} & \Nc{0.0800}
& \Sc{0.1324} & \Sc{0.1564} & \Nc{-0.0378} \\

2.4 Silence Elimination
& \Sc{0.0005} & \Sc{0.0023} & \Sc{0.0065}
& \Sc{0.0012} & \Nc{-0.0004} & \Nc{-0.0014}
& \Sc{0.0006} & \Tc{-0.0019} & \Nc{-0.0022}
& \Tc{-0.0006} & \Sc{0.0019} & \Nc{0.0066}
& \Nc{0.0000} & \Nc{0.0009} & \Nc{-0.0010} \\

\bottomrule
\end{tabular}
\end{adjustbox}
\label{tab:longitudinal-slope}
\end{table}

\begin{figure}[h]
  \centering
  \includegraphics{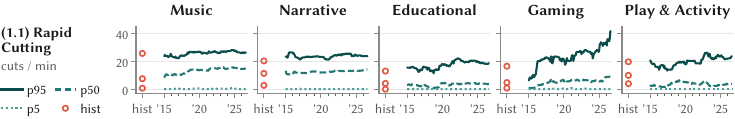}
  \vspace{-5pt}
  \caption{Longitudinal trend for cut rates. Cut rates increased significantly at one or more percentiles across all five content categories.}
  \label{fig:rapid-cutting-trend}
\end{figure}

\textbf{Cut rates increased across all categories of children's videos.} All categories showed a significant increase in (1.1) Rapid Cutting at one or more percentiles (Figure~\ref{fig:rapid-cutting-trend}). The 95th percentile increased in music, educational, gaming, and play \& activity videos, while the median increased in music, narrative, educational, and gaming videos. In contrast, the annual slopes at the 5th percentile were near zero in most categories.

\begin{figure}[h]
  \centering
  \includegraphics{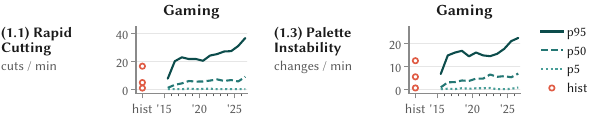}
  \vspace{-5pt}
  \caption{In gaming videos, the 95th percentile of (1.1) Rapid Cutting and (1.3) Palette Instability rose sharply, while the 5th stayed flat.}
  \label{fig:gaming-quantile-divergence}
\end{figure}

\textbf{The 95th percentiles often increased faster than others, while the 5th percentiles showed no detectable change.} In 12 of the 50 metric-category groups (24\%), the 95th percentile rose while the median and the 5th percentile increased more slowly, remained stable, or decreased. This pattern was most common in gaming and educational videos, each appearing on four metrics. For example, gaming videos' cut rates increased by 1.45 cuts/min per year at the 95th percentile, nearly three times the median rate of 0.50 cuts/min per year, while the 5th percentile remained unchanged (Figure~\ref{fig:gaming-quantile-divergence}). Overall, the 5th percentile changed significantly in only 10 groups (20\%), compared with 34 (68\%) at the median and 29 (58\%) at the 95th percentile.

\begin{figure}[t]
    \centering
    \includegraphics[width=\linewidth]{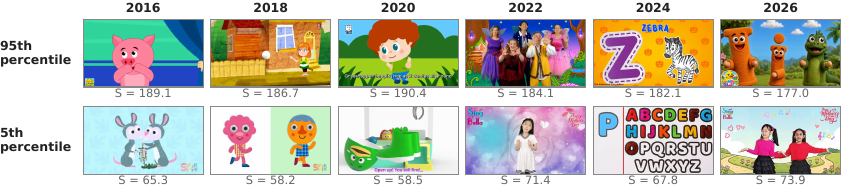}
    \caption{(2.1) Visual Intensity in music videos became more similar over time. From 2016 to 2026, saturation at the 95th percentile decreased from 189.1 to 177.0, and the 5th percentile increased from 65.3 to 73.9. Each image shows a frame from a video at the corresponding percentile, selected such that the frame's mean saturation is within 1 point of the percentile value.}
    \label{fig:saturation-trend-frames}
\end{figure}

\begin{figure}[h]
  \centering
  \includegraphics{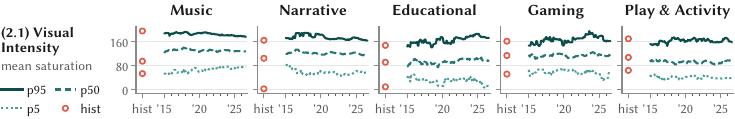}
  \vspace{-5pt}
  \caption{Longitudinal trend for (2.1) Visual Intensity. Saturation increased in educational and gaming videos, decreased in narrative videos, and became more similar across music videos.}
  \label{fig:saturation-trend}
\end{figure}

\begin{figure}[h]
  \centering
  \includegraphics{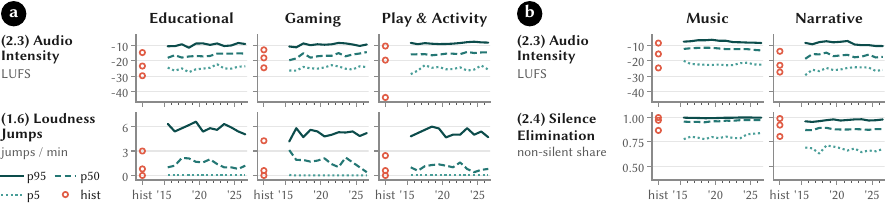}
  \begin{subfigure}{0pt}\phantomsubcaption\label{fig:loudness-educational-gaming-play}\end{subfigure}
  \begin{subfigure}{0pt}\phantomsubcaption\label{fig:loudness-music-narrative}\end{subfigure}
  \vspace{-15pt}
  \caption{Longitudinal trend for audio-related stimulation metrics.
    \textbf{(a)} Educational, gaming, and play \& activity videos became louder but had fewer sudden loudness jumps. \textbf{(b)} Music and narrative videos became quieter but had fewer silent moments.}
    \label{fig:loudness-trend}
\end{figure}

\textbf{Visual and audio stimulation followed contrasting trends across content categories.} Saturation increased in educational and gaming videos but decreased in narrative videos (Figure~\ref{fig:saturation-trend}). Music videos showed convergence, with saturation decreasing at the upper percentiles but increasing at the 5th percentile, narrowing the range by roughly 30 points over 12 years (Figure~\ref{fig:saturation-trend-frames}). Educational, gaming, and play \& activity videos became louder with fewer loudness jumps, while music and narrative videos became quieter with fewer silent moments (Figure~\ref{fig:loudness-trend}).

\begin{figure}[h]
  \centering
  \includegraphics{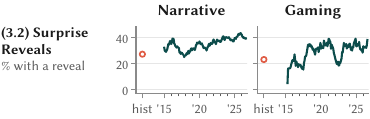}
  \caption{The proportion of narrative and gaming videos containing (3.2) Surprise Reveals increased over time.}
  \vspace{-5pt}
  \label{fig:surprise-reveal-trend}
\end{figure}

\textbf{Surprise Reveals became more prevalent in narrative and gaming videos.} Reward Patterning was relatively uncommon: 6.6\% of videos contained (3.1) Audio Build-Resolve events and 22.7\% contained (3.2) Surprise Reveals. The odds of containing a Surprise Reveal increased by 5.30\% per year in narrative videos and 5.34\% in gaming videos. Among videos exhibiting Surprise Reveals, their expected frequency also increased by 5.51\% per year in narrative videos, but not significantly in gaming videos (Figure~\ref{fig:surprise-reveal-trend}).

\subsection{RQ2: Are stimulation levels consistent within channels?}
\label{rq2}

\subsubsection{Method}
To assess within-channel consistency, we computed a one-way random-effects intraclass correlation coefficient~\cite{koo2016guideline} (ICC) for each of the 12 stimulation metrics, treating channel as a random effect. Higher ICC values indicate greater consistency among videos from the same channel. We restricted this analysis to channels with at least 10 sampled videos, retaining 494 channels and 16,206 videos.

To identify changes within individual channels over time, we further excluded channels whose sampled videos spanned less than one year, retaining 486 channels and 16,126 videos. For each metric, we fitted a random-intercept and random-slope model:
\[
\widehat{Y}_{m,i,j}
= (\beta_{0,m} + u_{0,m,i})
+ (\beta_{1,m} + u_{1,m,i}) \cdot \mathrm{Year}_{i,j},
\]
where $\widehat{Y}_{m,i,j}$ denotes the fitted value of stimulation metric \(m\) for video \(j\) from channel \(i\), and \(u_{0,m,i}\) and \(u_{1,m,i}\) represent channel-specific intercept and slope deviations. We identified significant channel-level increases and decreases using BH correction across all channel–metric tests ($p_{\mathrm{BH}}<.05$). We used 2022, the median video publication year, as a common reference point for comparing channels, since many channels had not yet published videos in earlier years.

\subsubsection{Results.}
\begin{figure}[h]
\centering
\footnotesize
\begin{minipage}[t]{12.2pt}
\vspace{0pt}\vspace{1.13pt}
\includegraphics{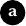}
\end{minipage}%
\hspace{4pt}%
\begin{minipage}[t]{0.21\linewidth}
\vspace{0pt}
\setlength{\tabcolsep}{3pt}
\renewcommand{\arraystretch}{1.12}
\begin{tabular}{@{}l r@{}}
\toprule
\textbf{Stimulation Pattern} & \textbf{ICC} \\
\midrule
1.1 Rapid Cutting          & \textbf{0.60} \\
1.2 Scene Discontinuity    & 0.38 \\
1.3 Palette Instability    & 0.44 \\
1.4 Luminance Flash        & 0.14 \\
1.5 Motion Density         & \textbf{0.53} \\
1.6 Loudness Jumps         & 0.39 \\
\addlinespace[2pt]
2.1 Visual Intensity       & 0.47 \\
2.2 Visual Brightness      & 0.41 \\
2.3 Audio Intensity        & \textbf{0.60} \\
2.4 Silence Elimination    & \textbf{0.50} \\
\addlinespace[2pt]
3.1 Audio Build-Resolve    & 0.03 \\
3.2 Surprise Reveals       & 0.21 \\
\bottomrule
\end{tabular}
\phantomsubcaption\label{fig:rq2-icc}
\end{minipage}%
\hfill
\begin{minipage}[t]{310.6pt}
\vspace{0pt}
\includegraphics{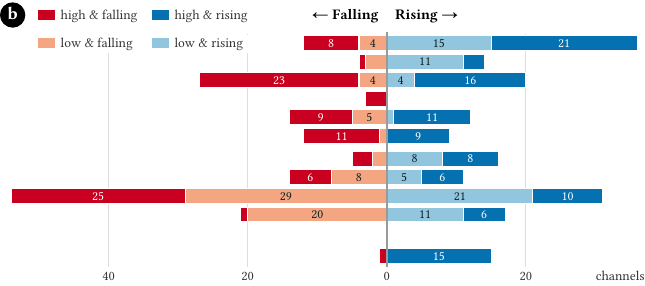}
\phantomsubcaption\label{fig:rq2-escalation}
\end{minipage}
\caption{
Within-channel consistency and longitudinal changes across
12 stimulation patterns.
(a)~ICC values, with bold indicating moderate consistency
($\geq 0.50$) and regular values indicating poor consistency
($<0.50$)~\cite{koo2016guideline}.
(b)~Numbers of channels with significant annual increases
or decreases ($p_{\mathrm{BH}}<.05$). Grouped by whether their estimated 2022 stimulation levels were above or below the median across channels.
}
\label{fig:rq2-summary}
\end{figure}

\textbf{Stimulation levels varied substantially within channels.} Eight of the 12 stimulation patterns had ICCs below 0.50, indicating poor within-channel consistency~\cite{koo2016guideline}. The remaining four showed moderate consistency ($\text{ICC} \geq 0.50$): (1.1) Rapid Cutting, (2.3) Audio Intensity, (1.5) Motion Density, and (2.4) Silence Elimination. We report the ICC estimates for all 12 stimulation patterns in Figure~\ref{fig:rq2-icc}.

\begin{figure}[h]
    \centering
    \includegraphics[width=0.5\linewidth]{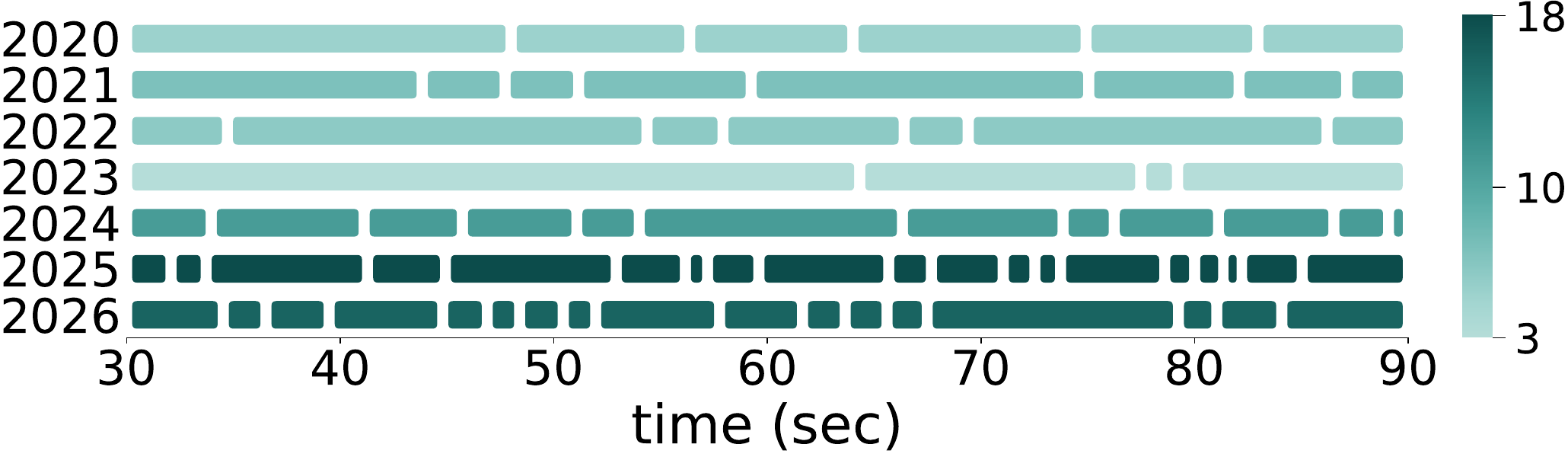}
    \vspace{-5pt}
    \caption{Shots became shorter over time in the \textit{Hop Hop the Owl} channel. Each row shows the 30--90 second segment of the earliest video in our dataset from each year. Blocks represent individual shots, with their widths indicating shot duration and boundaries indicating cuts. Color indicates the total number of cuts within the one-minute segment, with darker colors representing more cuts. 
    }
    \label{fig:cut-rate-hop-hop-the-owl}
\end{figure}

\textbf{Nearly a quarter of channels increased at least one stimulation pattern over time.} Overall, 116 channels (23.9\%) significantly increased at least one stimulation pattern. (1.1) Rapid Cutting showed the most widespread increase, with 36 channels (7.4\%) increasing and 12 (2.5\%) decreasing. For example, \textit{Hop Hop the Owl} increased its cut rate by 1.26 cuts/min per year ($p_{\mathrm{BH}}=0.014$), with annual mean cut rates rising from 6.5 in 2020 to 17.8 in 2025 (Figure~\ref{fig:cut-rate-hop-hop-the-owl}).

\textbf{Some patterns intensified primarily among channels with already high stimulation levels.} (1.1) Rapid Cutting increased among 21 channels above and 15 below the median baseline stimulation level. In contrast, 20 of the 21 channels with significant changes in (1.6) Loudness Jumps were above the median (Figure~\ref{fig:rq2-escalation}), while all 15 channels with increasing (3.2) Surprise Reveals were above the 75th percentile of estimated 2022 stimulation levels.

\subsection{RQ3: Do more stimulating videos receive more views?}
\label{rq3}

\subsubsection{Method}

To examine whether more stimulating videos receive more views, we fitted a linear mixed-effects model for each content category \(c\), using view counts as an indicator of video popularity:
\[
\widehat{\log(\mathrm{Views}_{i,j})}
= \alpha_c + \sum_{m=1}^{12}\gamma_{m,c}Y_{m,i,j}
+ \theta_c \cdot \mathrm{VideoAge}_{i,j} + u_{0,c,i},
\]
where \(Y_{m,i,j}\) denotes stimulation metric \(m\) for video \(j\) from channel \(i\), and \(u_{0,c,i}\) represents the channel-specific random intercept to account for differences in baseline popularity across channels. All five models had variance inflation factors (VIFs) below 3.0, indicating low multicollinearity among predictors. We log-transformed view counts to account for their heavy-tailed distribution~\cite{crane2008robust}. We also controlled for video age to account for differences in the time available to accumulate views. We converted coefficients to percentage changes in views using \((e^{\gamma_{m,c}}-1)\times100\%\) and applied BH correction across the 12 metric tests within each category.

\subsubsection{Results.}
\textbf{Specific stimulation patterns were associated with more views, with associations varying across content categories.} Of the 60 metric–category combinations, six showed significant positive associations, two showed negative associations, and 52 showed no significant associations after multiple-comparison correction (Table~\ref{tab:view-count-regression}). Note that these associations do not establish that stimulation causes videos to receive more views.

\begin{table}[h]
\centering
\footnotesize
\caption{Estimated percentage changes in video views per unit increase in each stimulation metric. Estimates are comparable across content categories within a metric, but not directly across metrics with different units. \colorbox{salmoncolor!45}{Salmon} and \colorbox{tealcolor!45}{teal} indicate significant positive and negative associations after BH correction. $^{*}p<0.05$, $^{**}p<0.01$, $^{***}p<0.001$.}
\label{tab:view-count-regression}

\setlength{\tabcolsep}{4pt}
\renewcommand{\arraystretch}{1}

\begin{tabular}{@{}l l l l l l@{}}
\toprule
\textbf{Metric} & \textbf{music} & \textbf{narrative} & \textbf{educational} & \textbf{gaming} & \textbf{play \& activity} \\
\midrule
\multicolumn{6}{@{}l}{\textbf{Pacing Intensification}} \\
1.1 Rapid Cutting (cuts / min)                   & \Nc{+0.66\%}   & \Sc{+3.14\%^{*\!*\!*}} & \Nc{+0.23\%}   & \Nc{+0.49\%}   & \Nc{+1.25\%} \\
1.2 Scene Discontinuity (mean inter-cut DSSIM)    & \Nc{-5.13\%}   & \Nc{-40.73\%}         & \Nc{-24.14\%}  & \Nc{-8.93\%}   & \Nc{-18.69\%} \\
1.3 Palette Instability (palette changes / min)   & \Nc{+0.76\%}   & \Nc{+0.76\%}          & \Nc{+1.70\%}   & \Nc{+1.24\%}   & \Nc{+3.05\%} \\
1.4 Luminance Flash (L* flashes / min)            & \Nc{-1.72\%}   & \Nc{-4.57\%}          & \Nc{+0.36\%}   & \Nc{-12.09\%}  & \Nc{-9.16\%} \\
1.5 Motion Density (fraction in motion)           & \Nc{+60.89\%}  & \Nc{-14.57\%}         & \Sc{+257.98\%^{*}} & \Nc{+53.24\%} & \Nc{-38.18\%} \\
1.6 Loudness Jumps (loudness jumps / min)         & \Nc{-2.07\%}   & \Nc{-0.26\%}          & \Nc{+1.83\%}   & \Nc{+4.71\%}   & \Tc{-6.90\%^{*\!*}} \\

\addlinespace[1pt]
\multicolumn{6}{@{}l}{\textbf{Sustained Sensory Intensity}} \\
2.1 Visual Intensity (mean saturation)            & \Nc{+0.06\%}   & \Nc{+0.06\%}          & \Nc{+0.03\%}   & \Nc{+0.24\%}   & \Sc{+0.60\%^{*\!*\!*}} \\
2.2 Visual Brightness (mean L*)                   & \Nc{-0.28\%}   & \Nc{-0.03\%}          & \Nc{+0.04\%}   & \Nc{+0.70\%}   & \Sc{+1.31\%^{*}} \\
2.3 Audio Intensity (integrated LUFS)             & \Nc{+1.58\%}   & \Nc{+0.17\%}          & \Nc{+2.53\%}   & \Sc{+7.18\%^{*\!*\!*}} & \Nc{+0.72\%} \\
2.4 Silence Elimination (fraction above the gate) & \Nc{+105.38\%} & \Tc{-67.24\%^{*\!*}}  & \Nc{+105.55\%} & \Nc{+139.02\%} & \Nc{+131.63\%} \\

\addlinespace[1pt]
\multicolumn{6}{@{}l}{\textbf{Reward Patterning}} \\
3.1 Audio Build-Resolve (build-resolves / min)    & \Sc{+145.72\%^{*}} & \Nc{+108.39\%} & \Nc{+6.65\%} & \Nc{+33.31\%} & \Nc{-17.27\%} \\
3.2 Surprise Reveals (reveals / min)              & \Nc{+6.55\%}   & \Nc{+6.04\%}          & \Nc{-26.49\%}  & \Nc{-7.54\%}   & \Nc{+0.23\%} \\
\midrule
$R^2$ conditional & 0.7013 & 0.6815 & 0.7529 & 0.6719 & 0.6920 \\
\bottomrule
\end{tabular}
\end{table}

\textbf{Videos with faster cutting or greater audio stimulation tended to receive more views in Narrative, Gaming, and Music categories.} In Narrative videos, faster cutting was positively associated with views (+3.14\% per additional cut/min), while greater silent moments were negatively associated. In Gaming videos, louder audio was associated with more views, while in Music videos, more frequent (3.1) Audio Build-Resolve events showed a positive association.

\textbf{Videos with greater visual stimulation tended to receive more views in Educational and Play \& Activity categories.} In Educational videos, a 10-percentage-point increase in (1.5) Motion Density was associated with 13.60\% more views. In Play \& Activity videos, (2.1) Visual Intensity (+0.60\% per unit) and (2.2) Visual Brightness (+1.31\% per unit) were positively associated with views, while (1.6) Loudness Jumps (-6.90\% per additional jump/min) were negatively associated.

\section{Public Facing Tool}
\label{sec:audit-tool}

\begin{figure*}[h]
  \centering
  \includegraphics[width=\textwidth]{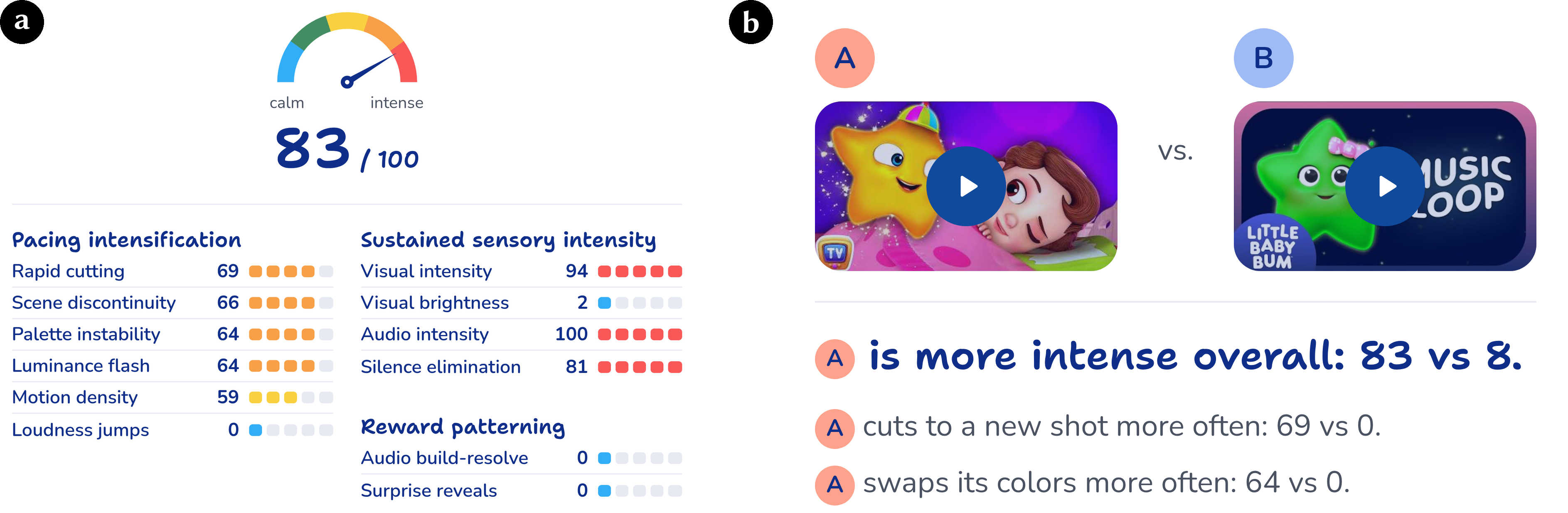}
  \caption{Web-based audit tool for exploring stimulation levels in child-facing videos. (a) Percentile-based scores and breakdowns across 12 hyperstimulation patterns. (b) Side-by-side comparison to help parents identify less stimulating alternatives.}
  \Description{Web-based audit tool for exploring stimulation levels in child-facing videos. (a) Percentile-based scores and breakdowns across 12 hyperstimulation patterns. (b) Side-by-side comparison to help parents identify less stimulating alternatives.}
  \label{fig:tool}
\end{figure*}

To help parents make informed viewing choices, we developed a web-based tool that makes stimulation levels in child-facing YouTube videos more transparent (Figure~\ref{fig:tool}). To make the measurements easier to interpret, we present stimulation levels as percentile scores on a 0–100 scale, with breakdowns across the 12 hyperstimulation patterns. Parents can compare videos to find less stimulating alternatives and explore channel rankings to identify potentially highly stimulating content. To keep the tool relevant as new videos emerge, a scheduled collection pipeline continuously discovers and analyzes new videos, while users can request analyses of videos not yet indexed. We plan to publicly release and open-source the tool, dataset, and measurement pipeline after publication.

\section{Discussion, Limitations, and Future Work}

\sssec{Data-Driven Optimization May Drive Hyperstimulation}.
Our audit found that stimulation in child-facing videos has increased over time, including within individual channels. 
One possible explanation is that this trend is a byproduct of data-driven optimization rather than a deliberate choice. 
Creators routinely use analytics dashboards, A/B testing of thumbnails and edits, and third-party optimization services that report how production choices relate to views and watch time~\cite{hill2022prominent, marshall2022abtesting}. 
A creator who notices that higher contrast, a louder mix, or a faster cut rate coincides with better performance has a clear incentive to repeat it; over many videos, such small, locally rewarded adjustments become habits, and a creator can arrive at a highly stimulating style without ever deciding to produce one~\cite{bishop2020algorithmic, macdonald2023actually}. When many creators follow the same incentives, the entire ecosystem may drift toward greater stimulation.

Creators cannot see this drift from inside their own analytics, which show how their videos perform but not how their production choices compare with peers or have changed over time. Our goal is to make this visible. Our public tool makes this visible by showing creators where their stimulation levels rank among similar channels and how their own channel has shifted. We hope this reference point helps creators make more deliberate and accountable decisions, and gives parents and regulators a basis for holding them to those decisions.

\sssec{Implications for Platforms}. Platforms are best positioned to break the cycle described above, because they control both the engagement signals creators optimize toward and the information parents see. Existing protections on child-facing video platforms focus on ensuring age-appropriate content~\cite{youtube-kids-best-practices, contentpolicies} and say little about how that content is produced. Platforms could surface stimulation indicators for individual videos, allow parents to filter videos by stimulation level, and show creators how a video's stimulation compares with similar videos before it is published. Such transparency would give parents more control and give creators a counterweight to engagement metrics when deciding how a video should look and sound.

\sssec{Limitations and Future Work}.
Our findings are bounded in three ways. First, we study YouTube only; children also watch TikTok, streaming services, and broadcast television, where production norms may differ, so our prevalence findings are specific to YouTube. Second, our keyword seeds produced an English-dominated corpus (73.2\% of videos declare English as their audio language), so our results may not generalize to other languages and markets. Third, our metrics characterize the video file rather than the viewing experience: a child watching on a phone at low volume is exposed to less stimulation than one watching on a television at high volume. Our measurements therefore support relative comparison across videos, not absolute stimulation levels for any individual child. Because our features are platform-agnostic, future work can extend this audit to other platforms and languages to test whether the upward trends we observe are specific to YouTube or reflect broader shifts in children's media.

\section{Conclusion}

We present a large-scale computational audit of hyperstimulation in 16,375 child-facing YouTube videos from 542 channels (2015–2026). Guided by parents' online discussions and developmental psychology, we operationalize 12 hyperstimulation patterns across three categories: pacing intensification, sustained sensory intensity, and reward patterning. We find increasing stimulation over time, with cut rates rising across all five content categories and nearly a quarter of channels showing increases in at least one pattern. Higher stimulation levels are also associated with more views, although the relevant patterns vary by content type. Finally, we develop a public lookup tool that allows parents to inspect and compare videos' stimulation levels. Together, our audit and tool make these production characteristics more transparent to support informed viewing choices.

\sssec{Generative AI Usage Statement}. 
We used ChatGPT (GPT-6) to polish the manuscript and Claude Code (Opus 5.5) to assist with writing code to produce the paper's figures. We also used Claude Sonnet 5 as an independent screener in a dual-screening process to identify relevant Reddit discussions (Appendix~\ref{app:coding}). The authors reviewed and validated AI-assisted outputs and take full responsibility for the content.

\sssec{Ethical Considerations Statement}. We followed established ethical guidelines for web scraping research~\cite{brown2025web, legewie2018youtube}. For the YouTube dataset, we collected only publicly accessible videos without circumventing access controls. We limited requests to 10 per minute to minimize server load and collected only the data necessary for our audit. For the Reddit analysis, we accessed data through the Reddit for Researchers program~\cite{reddit2026researchers}, which reviewed and approved our research plan. To respect copyright, we did not redistribute downloaded video clips and plan to release only video IDs in our open-source dataset. Our institution's IRB determined that the study did not constitute human subjects research and therefore did not require IRB approval.

\bibliographystyle{ACM-Reference-Format}
\bibliography{refs}

\appendix

\section{Channel Discovery Queries}
\label{app:search-queries}

Table~\ref{tab:search-queries} lists the search queries used to discover child-facing YouTube channels.

\begin{table}[!htbp]
\caption{Seed search queries for channel discovery.}
\centering
\footnotesize
\label{tab:search-queries}
\begin{tabular}{lp{0.85\linewidth}}
\toprule
\textbf{Category} & \textbf{Keyword Seeds} \\
\midrule
Music & nursery rhymes, kids songs, children's songs, sing along songs for kids, kids music videos, toddler songs \\
Narrative & cartoons for kids, kids animation series, stories for children, kids cartoon full episodes, storytime for kids, fairy tales for kids, preschool cartoon show \\
Educational & learning videos for kids, educational videos for children, kids science experiments, learn colors and shapes, phonics for kids, math for kids, stem for kids, preschool learning \\
Gaming & kids gaming, minecraft for kids, roblox gameplay for kids, family friendly gaming, kids game commentary, video games for kids \\
Play \& activity & toy play for kids, kids crafts, diy for kids, kids challenge videos, play doh videos for kids, arts and crafts for children, sensory play for toddlers, outdoor activities for kids, kids exercise \\
\bottomrule
\end{tabular}
\end{table}

\section{Contemporary YouTube Corpus Distribution}
\label{app:contemporary-distribution}
\begin{figure}[h]
    \centering
    \includegraphics[width=0.5\linewidth]{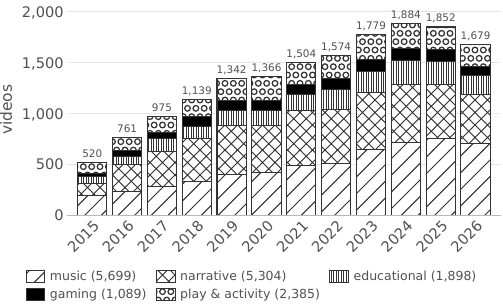}
    \vspace{-5pt}
    \caption{Distribution of contemporary YouTube videos by publication year and content category.}
    \label{fig:contemporary-distribution}
\end{figure}

Figure~\ref{fig:contemporary-distribution} shows the distribution of our YouTube video dataset by year and content category.

\section{Contemporary Video Category Classification}
\label{app:video-classification}

Table~\ref{tab:classification-keywords} lists the keywords used to classify videos into five content categories.

\begin{table}[!htbp]
\centering
\footnotesize
\caption{Keyword lists used for contemporary video category classification.}
\label{tab:classification-keywords}
\begin{tabular}{l p{0.6\linewidth}}
\toprule
\textbf{Category} & \textbf{Keywords} \\
\midrule
Music &
kidsong, lyrics, music, nursery rhyme, sing along, song \\
\addlinespace
Narrative &
cartoon, fairy tale, full episode, read-aloud, story \\
\addlinespace
Educational &
education, experiment, grammar, history, knowledge, learn, lesson, math, phonics, science, STEM, study, vocabulary \\
\addlinespace
Gaming &
game, gameplay, gaming, minecraft, nintendo, racing, roblox \\
\addlinespace
Play \& Activity &
challenge, craft, dance, DIY, draw, exercise, fitness, lego, painting, play, swim, toy, vlog, workout, yoga, zumba \\
\bottomrule
\end{tabular}
\end{table}

\section{Historical Reference Set Curation}
\label{app:historical-dataset}

We collected historical children’s videos originally uploaded to YouTube before 2015, along with older television content re-uploaded by official or fan archive channels. We determined each video's original production year from its air date or title information and assigned it to one of the five content categories used for the contemporary corpus. Table~\ref{tab:historical-corpus} summarizes the five production eras and their representative content.

\begin{table}[!htbp]
\footnotesize
\caption{Historical reference set by production era.}
\label{tab:historical-corpus}
\begin{tabular}{@{}llrp{9cm}@{}}
\toprule
Era & Years & Videos & Examples \\
\midrule
E0: Theatrical animation & 1928--1946 & 30 & Mickey Mouse, Silly Symphonies, Betty Boop, Popeye, Mighty Mouse \\
E1: Early television & 1947--1959 & 28 & Looney Tunes, Woody Woodpecker, Colonel Bleep, Watch Mr.\ Wizard, Ding Dong School \\
E2: Network television & 1960--1979 & 52 & Sesame Street, Mister Rogers' Neighborhood, Schoolhouse Rock, Pink Panther, ZOOM \\
E3: Cable television & 1980--1999 & 85 & Barney \& Friends, Care Bears, Garfield and Friends, Nickelodeon GUTS, Art Attack, Starcade \\
E4: Early online video & 2000--2014 & 105 & Zoboomafoo, Yo Gabba Gabba!, Boohbah, LittleBabyBum, ChuChu TV, Game Grumps \\
\midrule
Total & & 300 & \\
\bottomrule
\end{tabular}
\end{table}

\FloatBarrier

\section{Subreddits}
\label{app:subreddits}

Table~\ref{tab:subreddits} lists the eight subreddits included in our analysis.

\begin{table}[!htbp]
\centering
\footnotesize
\caption{We selected 8 subreddits to understand parents' concerns about hyper-stimulating videos.}
\begin{tabular}{lp{0.63\linewidth}l}
\toprule
\textbf{Subreddit} & \textbf{Description} & \textbf{Visits/wk} \\
\midrule
r/Parenting & For those with kids of any age & 3.3M \\
r/NewParents & A place to share thoughts, questions, support, and tips about being a new parent & 2.3M \\
r/daddit & A community for dads & 1.1M \\
r/toddlers & A community for parents and caregivers of kids ages 1--3 & 751K \\
r/Mommit & A support community for moms navigating the messier parts of child-rearing & 740K \\
r/beyondthebump & A community for new parents transitioning into life with their little one & 729K \\
r/ScienceBasedParenting & A place to discuss parenting questions and research through a science-based lens & 395K \\
r/preschoolers & A place for parents and caregivers of preschoolers (ages 3--5) to share stories and seek advice & 123K \\
\bottomrule
\end{tabular}
\label{tab:subreddits}
\end{table}

\FloatBarrier

\section{Qualitative Coding Procedure and Codebook}
\label{app:coding}

One author and an LLM (Claude Sonnet 5) independently screened the collected Reddit discussions for relevance, achieving Gwet's AC1 of 0.86 for posts and 0.90 for comments. Two authors then independently coded an initial 5\% sample and refined the codebook through three additional 5\% samples until reaching saturation. The resulting codebook covered two themes: (1) concerning video characteristics and (2) perceived effects of hyperstimulation. The authors then independently coded another 10\% of the data, achieving Gwet's AC1 of 0.93, resolved disagreements, and divided the remaining data for coding. Table~\ref{tab:codebook} presents the final codebook.

\begin{table}[!htbp]
\caption{Codebook for parents' Reddit discussions of hyperstimulation.}
\footnotesize
\label{tab:codebook}
\begin{tabular}{@{}p{2.0cm}p{3.5cm}p{8cm}@{}}
\toprule
\textbf{Theme} & \textbf{Code} & \textbf{Definition} \\
\midrule
\multirow{7}{2.6cm}{Concerning video characteristics}
& Fast cuts & Parent mentions rapid scene/shot changes or quick edits. \\
\cmidrule(l){2-3}
& Bright colors & Parent describes the video as visually bright. \\
\cmidrule(l){2-3}
& Highly saturated colors & Parent references intense, oversaturated, vivid colors. \\
\cmidrule(l){2-3}
& Color changing & Parent describes the video's colors or palette constantly shifting or changing. \\
\cmidrule(l){2-3}
& Loud & Parent flags the video's excessive audio volume. \\
\cmidrule(l){2-3}
& Silence elimination & Parent notes noisy, or never a quiet moment. \\
\cmidrule(l){2-3}
& Flashing visual effect & Parent describes flickering, or sudden flashes of bright colors. \\
\cmidrule(l){2-3}
& Lots of motion or camera movement & Parent describes constant camera movement or objects in continuous motion. \\
\midrule
\multirow{5}{2.6cm}{Perceived effects of hyperstimulation}
& Tantrum & Child has an acute meltdown/tantrum, often specifically triggered by the video ending or being taken away (includes refusing to turn off). \\
\cmidrule(l){2-3}
& Shortened attention span & Parent perceives a reduction in the child's ability to focus after watching. \\
\cmidrule(l){2-3}
& Zone out & Child appears unresponsive while watching. \\
\cmidrule(l){2-3}
& Bad tempered & Child becomes irritable or short-tempered after watching. \\
\cmidrule(l){2-3}
& Hyperactive & Parent describes the child unable to calm down after watching. \\
\bottomrule
\end{tabular}
\end{table}

\section{Additional Details on Statistical Methods}

\subsection{RQ1: Statistical Models and Diagnostics}
\label{app:rq1-methods}

For the quantile regressions, we calculated category-specific
annual slopes by combining the Year coefficient with the
corresponding Year $\times$ Category interaction coefficient:
\begin{equation}
    \text{Slope}_{c,\tau}
    = \beta_1^\tau + \beta_{3,c}^\tau,
\end{equation}
where $\beta_{3,c}^\tau = 0$ for the reference category.
We assessed category-specific trends using 95\% confidence
intervals and tested the overall Year $\times$ Category
interaction to determine whether temporal trends differed
across content categories.

For the two Reward Patterning metrics, we specified the
occurrence component of the hurdle model as
\begin{equation}
    \operatorname{logit}(P(Y_m > 0))
    = \beta_0 + \beta_1 \cdot \text{Year}
    + \beta_2 \cdot \text{Category}
    + \beta_3 \cdot (\text{Year} \times \text{Category}),
\end{equation}
and the positive-count component as
\begin{equation}
    \log(E[Y_m \mid Y_m > 0])
    = \gamma_0 + \gamma_1 \cdot \text{Year}
    + \gamma_2 \cdot \text{Category}
    + \gamma_3 \cdot (\text{Year} \times \text{Category}).
\end{equation}

We exponentiated the category-specific Year coefficients
to obtain odds ratios for the occurrence component and
rate ratios for the positive-count component. These ratios
represent multiplicative annual changes in the odds of
observing a pattern and its expected frequency among
videos exhibiting the pattern, respectively.

\subsection{RQ2: Channel-Level Model Details}
\label{app:rq2-methods}

We used the one-way, single-measurement random-effects ICC, calculated as:
\begin{equation}
    \mathrm{ICC}(1,1) =
    \frac{\sigma^2_{\mathrm{Channel}}}
    {\sigma^2_{\mathrm{Channel}} + \sigma^2_{\mathrm{Residual}}},
\end{equation}
where $\sigma^2_{\mathrm{Channel}}$ and
$\sigma^2_{\mathrm{Residual}}$ represent between-channel
and within-channel variance, respectively. For the channel-level trend analysis, we obtained each channel's annual slope by combining the fixed Year coefficient with its channel-specific random slope:
\begin{equation}
\mathrm{Slope}_{m,i} = \beta_{1,m} + u_{1,m,i},
\end{equation}
where $m$ denotes the stimulation metric and $i$ denotes the channel.

\section{Full Model Estimates for RQ1}
\label{app:model-estimates-rq1}

Table~\ref{tab:longitudinal-slope-app} and Table~\ref{tab:or-rr-table-app} report the full quantile regression and hurdle model estimates, respectively.

\begin{table}[!htbp]
\centering
\scriptsize
\caption{Slope estimates. $p$ indicates BH-corrected significance of slope differences across categories. Cells are shaded when the 95\% CI excludes zero. \colorbox{salmoncolor!45}{Salmon} indicates positive slopes and \colorbox{tealcolor!45}{teal} negative slopes.
}
\renewcommand{\arraystretch}{1.15}
\setlength{\tabcolsep}{3pt}
\resizebox{\linewidth}{!}{%
\begin{tabular}{p{\metriccolwidth} c c *{5}{c}}
\toprule
\multirow{2}{*}{Metric} & \multirow{2}{*}{$\tau$} & \multirow{2}{*}{$p$}
 & \multicolumn{5}{c}{slope/yr\ \ [95\% CI]} \\
 & & & music & narrative & educational & gaming & play \& activity \\
\midrule
 
\multirow{3}{\metriccolwidth}{1.1 Rapid Cutting (cuts/min)}
 & p95  & ***  & \Sc{0.1550}{[0.0152,\ 0.2948]} & \Nc{0.1330}{[-0.0215,\ 0.2875]} & \Sc{0.5204}{[0.2881,\ 0.7527]} & \Sc{1.4541}{[1.0839,\ 1.8242]} & \Sc{0.6898}{[0.4666,\ 0.9131]} \\
 & p50  & ***  & \Sc{0.4026}{[0.3272,\ 0.4781]} & \Sc{0.2389}{[0.1580,\ 0.3199]} & \Sc{0.2316}{[0.1012,\ 0.3620]} & \Sc{0.5003}{[0.3243,\ 0.6763]} & \Nc{-0.0075}{[-0.1230,\ 0.1079]} \\
 & p5   & -     & \Sc{0.0211}{[0.0003,\ 0.0419]} & \Nc{0.0000}{[-0.0275,\ 0.0275]} & \Nc{0.0000}{[-0.0636,\ 0.0636]} & \Nc{0.0000}{[-0.0640,\ 0.0640]} & \Nc{0.0000}{[-0.0543,\ 0.0543]} \\
\midrule
 
\multirow{3}{\metriccolwidth}{1.2 Scene Discontinuity (DSSIM)}
 & p95  & -    & \Tc{-0.0026}{[-0.0039,\ -0.0013]} & \Nc{-0.0008}{[-0.0022,\ 0.0006]} & \Nc{-0.0015}{[-0.0036,\ 0.0006]} & \Nc{0.0010}{[-0.0019,\ 0.0039]} & \Nc{-0.0010}{[-0.0031,\ 0.0010]} \\
 & p50  & ***  & \Sc{0.0036}{[0.0025,\ 0.0046]} & \Sc{0.0013}{[0.0002,\ 0.0024]} & \Sc{0.0047}{[0.0028,\ 0.0065]} & \Sc{0.0079}{[0.0055,\ 0.0102]} & \Tc{-0.0031}{[-0.0047,\ -0.0015]} \\
 & p5   & ***  & \Sc{0.0104}{[0.0072,\ 0.0137]} & \Nc{-0.0006}{[-0.0037,\ 0.0025]} & \Nc{0.0024}{[-0.0031,\ 0.0080]} & \Nc{0.0045}{[-0.0038,\ 0.0128]} & \Nc{-0.0036}{[-0.0086,\ 0.0014]} \\
\midrule
 
\multirow{3}{\metriccolwidth}{1.3 Palette Instability (changes/min)}
 & p95  & ***  & \Tc{-0.3485}{[-0.4726,\ -0.2245]} & \Nc{0.0000}{[-0.1465,\ 0.1465]} & \Sc{0.2880}{[0.0575,\ 0.5184]} & \Sc{0.7091}{[0.3928,\ 1.0254]} & \Nc{0.0051}{[-0.1832,\ 0.1934]} \\
 & p50  & ***  & \Tc{-0.1165}{[-0.1688,\ -0.0643]} & \Sc{0.1040}{[0.0480,\ 0.1600]} & \Nc{0.0872}{[-0.0030,\ 0.1774]} & \Sc{0.3874}{[0.2657,\ 0.5092]} & \Nc{0.0256}{[-0.0542,\ 0.1055]} \\
 & p5   & -     & \Nc{-0.0067}{[-0.0293,\ 0.0158]} & \Nc{0.0000}{[-0.0219,\ 0.0219]} & \Nc{0.0000}{[-0.0434,\ 0.0434]} & \Nc{0.0339}{[-0.0203,\ 0.0880]} & \Nc{0.0000}{[-0.0464,\ 0.0464]} \\
\midrule
 
\multirow{3}{\metriccolwidth}{1.4 Luminance Flash (flashes/min)}
 & p95  & ***  & \Tc{-0.0358}{[-0.0545,\ -0.0172]} & \Nc{-0.0003}{[-0.0213,\ 0.0207]} & \Nc{-0.0027}{[-0.0357,\ 0.0303]} & \Sc{0.0845}{[0.0444,\ 0.1246]} & \Nc{-0.0194}{[-0.0459,\ 0.0072]} \\
 & p50  & -     & \Nc{0.0000}{[-0.0015,\ 0.0015]} & \Nc{0.0000}{[-0.0016,\ 0.0016]} & \Nc{0.0000}{[-0.0025,\ 0.0025]} & \Nc{0.0000}{[-0.0034,\ 0.0034]} & \Nc{0.0000}{[-0.0022,\ 0.0022]} \\
 & p5   & -     & \Nc{0.0000}{[-0.0007,\ 0.0007]} & \Nc{0.0000}{[-0.0007,\ 0.0007]} & \Nc{0.0000}{[-0.0015,\ 0.0015]} & \Nc{0.0000}{[-0.0017,\ 0.0017]} & \Nc{0.0000}{[-0.0014,\ 0.0014]} \\
\midrule
 
\multirow{3}{\metriccolwidth}{1.5 Motion Density (fraction)}
 & p95  & ***  & \Nc{-0.0032}{[-0.0067,\ 0.0004]} & \Nc{0.0035}{[-0.0007,\ 0.0077]} & \Nc{-0.0042}{[-0.0106,\ 0.0022]} & \Tc{-0.0238}{[-0.0327,\ -0.0150]} & \Tc{-0.0064}{[-0.0121,\ -0.0008]} \\
 & p50  & ***  & \Sc{0.0061}{[0.0048,\ 0.0075]} & \Sc{0.0026}{[0.0011,\ 0.0040]} & \Sc{0.0039}{[0.0016,\ 0.0061]} & \Nc{0.0020}{[-0.0011,\ 0.0050]} & \Tc{-0.0079}{[-0.0099,\ -0.0058]} \\
 & p5   & ***  & \Sc{0.0020}{[0.0015,\ 0.0025]} & \Nc{0.0000}{[-0.0006,\ 0.0005]} & \Nc{-0.0003}{[-0.0012,\ 0.0007]} & \Nc{0.0007}{[-0.0007,\ 0.0020]} & \Tc{-0.0008}{[-0.0015,\ -0.0001]} \\
\midrule
 
\multirow{3}{\metriccolwidth}{1.6 Loudness Jumps (jumps/min)}
 & p95  & *    & \Tc{-0.0844}{[-0.1222,\ -0.0467]} & \Nc{0.0001}{[-0.0416,\ 0.0417]} & \Tc{-0.1056}{[-0.1711,\ -0.0401]} & \Nc{-0.0311}{[-0.1218,\ 0.0596]} & \Nc{-0.0372}{[-0.0960,\ 0.0216]} \\
 & p50  & ***  & \Nc{0.0000}{[-0.0051,\ 0.0051]} & \Sc{0.0617}{[0.0563,\ 0.0672]} & \Tc{-0.0731}{[-0.0819,\ -0.0643]} & \Tc{-0.1166}{[-0.1285,\ -0.1047]} & \Tc{-0.0515}{[-0.0593,\ -0.0436]} \\
 & p5   & -     & \Nc{0.0000}{[-0.0055,\ 0.0055]} & \Nc{0.0000}{[-0.0035,\ 0.0035]} & \Nc{0.0000}{[-0.0058,\ 0.0058]} & \Nc{0.0000}{[-0.0075,\ 0.0075]} & \Nc{0.0000}{[-0.0060,\ 0.0060]} \\
\midrule
 
\multirow{3}{\metriccolwidth}{2.1 Visual Intensity (S)}
 & p95  & ***  & \Tc{-1.3156}{[-1.9079,\ -0.7232]} & \Tc{-1.6730}{[-2.3459,\ -1.0000]} & \Sc{1.8950}{[0.8519,\ 2.9381]} & \Sc{1.7204}{[0.3401,\ 3.1006]} & \Nc{0.4196}{[-0.4688,\ 1.3081]} \\
 & p50  & ***  & \Tc{-0.5351}{[-0.8988,\ -0.1713]} & \Tc{-0.6875}{[-1.0774,\ -0.2975]} & \Sc{1.2948}{[0.6663,\ 1.9232]} & \Nc{0.3371}{[-0.5112,\ 1.1854]} & \Nc{-0.0916}{[-0.6478,\ 0.4647]} \\
 & p5   & ***  & \Sc{1.7950}{[1.1284,\ 2.4615]} & \Nc{-0.1246}{[-0.7813,\ 0.5320]} & \Nc{-0.6319}{[-1.8106,\ 0.5468]} & \Nc{-1.3163}{[-2.9403,\ 0.3078]} & \Nc{-0.2823}{[-1.2852,\ 0.7206]} \\
\midrule
 
\multirow{3}{\metriccolwidth}{2.2 Visual Brightness (L*)}
 & p95  & ***  & \Tc{-0.6091}{[-0.8171,\ -0.4012]} & \Sc{0.4192}{[0.2147,\ 0.6237]} & \Nc{0.1533}{[-0.2239,\ 0.5306]} & \Sc{0.7538}{[0.2713,\ 1.2362]} & \Sc{0.4198}{[0.1116,\ 0.7279]} \\
 & p50  & ***  & \Sc{0.1548}{[0.0284,\ 0.2812]} & \Sc{0.1888}{[0.0533,\ 0.3243]} & \Nc{0.0395}{[-0.1789,\ 0.2579]} & \Sc{0.8744}{[0.5796,\ 1.1691]} & \Nc{0.1830}{[-0.0102,\ 0.3763]} \\
 & p5   & **   & \Sc{0.6121}{[0.3116,\ 0.9125]} & \Nc{0.1006}{[-0.2636,\ 0.4648]} & \Sc{0.5738}{[0.0306,\ 1.1171]} & \Nc{0.5422}{[-0.1602,\ 1.2445]} & \Tc{-0.5645}{[-1.0955,\ -0.0335]} \\
\midrule
 
\multirow{3}{\metriccolwidth}{2.3 Audio Intensity (LUFS)}
 & p95  & ***  & \Tc{-0.1996}{[-0.2479,\ -0.1514]} & \Tc{-0.3693}{[-0.4164,\ -0.3221]} & \Nc{0.0674}{[-0.0188,\ 0.1535]} & \Nc{0.0256}{[-0.0881,\ 0.1393]} & \Sc{0.1324}{[0.0501,\ 0.2148]} \\
 & p50  & ***  & \Tc{-0.1333}{[-0.1858,\ -0.0807]} & \Tc{-0.1426}{[-0.1990,\ -0.0862]} & \Sc{0.2775}{[0.1867,\ 0.3684]} & \Sc{0.2545}{[0.1319,\ 0.3772]} & \Sc{0.1564}{[0.0759,\ 0.2369]} \\
 & p5   & -    & \Nc{0.0047}{[-0.0820,\ 0.0915]} & \Nc{0.0560}{[-0.0556,\ 0.1676]} & \Sc{0.2675}{[0.1115,\ 0.4235]} & \Nc{0.0800}{[-0.1479,\ 0.3078]} & \Nc{-0.0378}{[-0.1870,\ 0.1113]} \\
\midrule
 
\multirow{3}{\metriccolwidth}{2.4 Silence Elimination (fraction)}
 & p95  & ***  & \Sc{0.0005}{[0.0002,\ 0.0007]} & \Sc{0.0012}{[0.0009,\ 0.0015]} & \Sc{0.0006}{[0.0001,\ 0.0011]} & \Tc{-0.0006}{[-0.0011,\ 0.0000]} & \Nc{0.0000}{[-0.0004,\ 0.0005]} \\
 & p50  & ***  & \Sc{0.0023}{[0.0016,\ 0.0029]} & \Nc{-0.0004}{[-0.0011,\ 0.0002]} & \Tc{-0.0019}{[-0.0030,\ -0.0008]} & \Sc{0.0019}{[0.0004,\ 0.0034]} & \Nc{0.0009}{[0.0000,\ 0.0019]} \\
 & p5   & *    & \Sc{0.0065}{[0.0028,\ 0.0101]} & \Nc{-0.0014}{[-0.0054,\ 0.0025]} & \Nc{-0.0022}{[-0.0086,\ 0.0041]} & \Nc{0.0066}{[-0.0027,\ 0.0159]} & \Nc{-0.0010}{[-0.0071,\ 0.0050]} \\
 
\bottomrule
\multicolumn{8}{l}{\footnotesize $-$: $p\geq0.05$, *: $p<0.05$, **: $p<0.01$, ***: $p<0.001$} \\
\end{tabular}%
}
\label{tab:longitudinal-slope-app}
\end{table}

\begin{table}[!htbp]
\centering
\scriptsize
\caption{Ratio estimates. $p$ indicates BH-corrected significance across categories. Cells are shaded when the 95\% CI excludes 1. \colorbox{salmoncolor!45}{Salmon} indicates significant increases and \colorbox{tealcolor!45}{teal} decreases. Odds ratios represent annual changes in occurrence odds ($P(Y_m)>0$), while rate ratios represent annual changes in expected counts among videos with occurrences ($E[Y_m \mid Y_m>0]$). For (3.2) Surprise Reveals in narrative videos, odds and rate ratios of 1.0530 and 1.0551 indicate annual increases of 5.3\% and 5.5\%, respectively.
}
\renewcommand{\arraystretch}{1.15}
\setlength{\tabcolsep}{3pt}
\resizebox{\linewidth}{!}{%
\begin{tabular}{p{\metriccolwidth} c c *{5}{c}}
\toprule
\multirow{2}{*}{Metric} & \multirow{2}{*}{part} & \multirow{2}{*}{$p$}
 & \multicolumn{5}{c}{Ratio \ [95\% CI]} \\
 & & & music & narrative & educational & gaming & play \& activity \\
\midrule

\multirow{2}{\metriccolwidth}{3.1 Audio Build-Resolve}
 & $P(Y_m) > 0$  & * & \Nc{1.0277}{[0.9915,\ 1.0652]} & \Nc{0.9928}{[0.9607,\ 1.0259]} & \Tc{0.9007}{[0.8426,\ 0.9628]} & \Nc{0.9923}{[0.9258,\ 1.0636]} & \Nc{1.0017}{[0.9517,\ 1.0543]} \\
 & $E[Y_m \mid Y_m > 0]$  & -  & \Nc{0.9714}{[0.8817,\ 1.0702]} & \Nc{1.0562}{[0.9423,\ 1.1837]} & \Nc{1.0111}{[0.8752,\ 1.1682]} & \Nc{0.9326}{[0.7660,\ 1.1356]} & \Nc{1.0376}{[0.8551,\ 1.2592]} \\
\midrule

\multirow{2}{\metriccolwidth}{3.2 Surprise Reveals}
 & $P(Y_m) > 0$  & -  & \Nc{1.0093}{[0.9824,\ 1.0368]} & \Sc{1.0530}{[1.0335,\ 1.0728]} & \Nc{1.0365}{[0.9986,\ 1.0758]} & \Sc{1.0534}{[1.0094,\ 1.0992]} & \Nc{1.0227}{[0.9924,\ 1.0539]} \\
 & $E[Y_m \mid Y_m > 0]$  & -  & \Nc{0.9912}{[0.9501,\ 1.0340]} & \Sc{1.0551}{[1.0300,\ 1.0807]} & \Nc{1.0568}{[0.9942,\ 1.1232]} & \Nc{1.0292}{[0.9676,\ 1.0948]} & \Sc{1.0481}{[1.0031,\ 1.0951]} \\

\bottomrule
\multicolumn{8}{l}{\footnotesize -: $p \geq 0.05$, *: $p<0.05$} \\
\end{tabular}%
}
\label{tab:or-rr-table-app}
\end{table}

\FloatBarrier

\section{Longitudinal Trend of Stimulation Patterns}
\label{app:longitudinal-trend}

\begin{figure}[!htbp]
  \centering
  \includegraphics{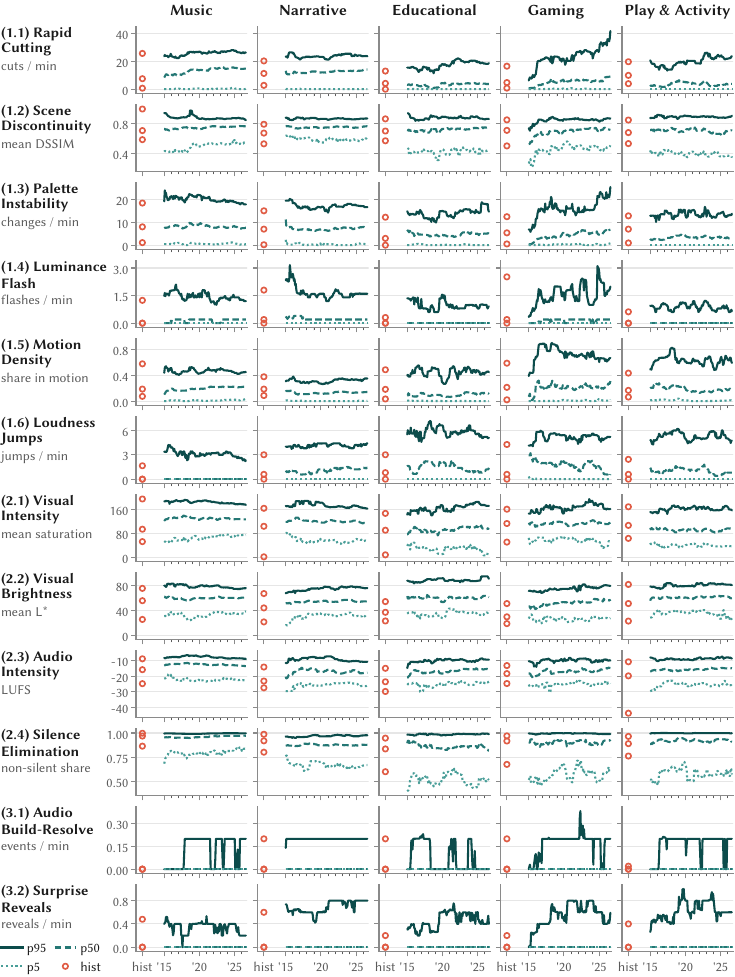}
  \caption{Longitudinal trend of all 12 stimulation patterns.}
  \label{fig:longitudinal-trend-all}
\end{figure}

Figure~\ref{fig:longitudinal-trend-all} presents the longitudinal trends for all 12 stimulation patterns.

\end{document}